\documentclass[showpacs,superscriptaddress,floatfix,pra,twocolumn,10pt]{revtex4}
\usepackage{graphicx}
\usepackage{amsmath,amssymb}
\usepackage{bm}

\input{tcilatex}

\begin{document}

\title{Nonlinearity enhancement in optomechnical system}
\author{Ling Zhou\thanks{%
Email: Zhlhxn@dlut.edu.cn}}
\affiliation{School of physics and optoelectronic technology, Dalian University of
Technology, Dalian 116024, P.R.China}
\author{Jiong Cheng}
\affiliation{School of physics and optoelectronic technology, Dalian University of
Technology, Dalian 116024, P.R.China}
\author{Yan Han}
\affiliation{School of physics and optoelectronic technology, Dalian University of
Technology, Dalian 116024, P.R.China}
\author{Weiping Zhang}
\affiliation{State Key Laboratory of Precision Spectroscopy, Department of Physics, East
China Normal University, Shanghai 200062, P.R. China}

\begin{abstract}
The nonlinearity is an important feature in the field of optomechanics.
Employing atomic coherence, we put forward a scheme to enhance the
nonlinearity of the cavity optomechanical system. The effective Hamiltonian
is derived, which shows that the nonlinear strength can be enhanced by
increasing the number of atoms at certain range of parameters. We also
numerically study the nonlinearity enhancement beyond the effective
Hamiltonian. Furthermore, we investigate the potential usage of the
nonlinearity in performing quantum nondemolition (QND) measurement of the
bosonic modes. Our results show that the present system exhibits
synchronization, and the nonlinear effects provide us means in performing
QND.
\end{abstract}

\maketitle

\baselineskip=18pt

\section{\protect\bigskip\ Introduction}

Optomechanical system by coupling mechanical resonators to the fields of
optical cavities can provide us available device to observe quantum
mechanical behaviors of macroscopic system. It has been proved that
entanglement of two resonators and of cavity and mirror can be generated by
radiation pressure \cite{entm,entm2,zhang,zhou}, while the pressure can be
used to cool down the mirrors \cite{cooling,cooling2,liy,cooling3,cooling4},
and the analogy electromagnetically induced transparency phenomenon may
happen in cavity optomechanical system \cite{eit,hanyan,hanyan2,qiteng} and
has demonstrated in experiment \cite{eitex}. In addition, optomechanical
system can be used as transducers for long-distance quantum communication %
\cite{transducer1,transducer2}.

On the other hand, the observation of strict quantum effects in quantum
optics relies on the existence of strong nonlinear interaction between
photons \cite{imam}. Unfortunately, photons tend to interact only weakly,
thus enhancement of photon-photon interaction at the few-photon level is
still a challenge in quantum optics. A lot of efforts are devoted to enhance
the nonlinearity of photons \cite{imam2,gen1,hart,zhounon}. Gong et. al. %
\cite{gongzr} has shown us that an optomechanical system can lead to
nonlinear Kerr effect, but it is very weak ( proportion to $\frac{G^{2}}{%
\omega _{m}^{2}}$) because usually the radiation pressure coupling strength $%
G$ is less than the frequency of the oscillator $\omega _{m}$ for weak
coupling system. Most recently, Ludwig et. al. \cite{max} propose a scheme
to enhance cross-Kerr nonlinearity in double cavities with membrane in the
middle where the tunnel rate between cavities weaken the negative influence
of large value of the oscillator frequency.

Optomechanics experiments are rapidly approaching the single-photon
strong-coupling regime $G\geq \omega _{m}$ \cite{Num}. However, for weakly
driven systems \cite{rabl,Num}, Rabl \cite{rabl} has shown that photon
blockade under single-photon strong coupling condition is affected by $\frac{%
G}{\omega _{m}}$ where for $\frac{G}{\omega _{m}}>\frac{1}{2}$, no
significant further improvement of the nonlinearity is achieved, and the
nonlinear effects are suppressed by $\frac{G}{\omega _{m}}$. Thus, improving
the nonlinearity beyond strong-coupling means deserves our investigation. In
this paper, we consider weakly driven system with weak coupling $G<\omega
_{m}$ and $G^{2}<\kappa \omega _{m}$ and put forward an alternative scheme
to enhance and to modulate the photon-photon and photon-phonon cross-Kerr
nonlinearity by employing atomic coherence. We also show the nondemolition
measurement of phonon and photon. Comparing the scheme with \cite%
{max,rabl,Num}, the photon-photon and photon-phonon cross-Kerr
nonlinearities not only can be enhanced but also can be controlled. We do
not need single-photon strong coupling condition, which means that our
scheme maybe easier to be realized.

\bigskip

\section{MODEL AND effective interaction}

We consider atomic media trapped in a doubly resonant cavity with one
partially transmitting fixed mirror and one movable mirror (see Fig. 1). The
two cavity modes with frequencies $\omega _{1}$ and $\omega _{2}$ couple to
atomic transition $|a\rangle \leftrightarrow |c\rangle $ and $|a\rangle
\leftrightarrow |b\rangle $ with the same detuning $\delta $, and the
classical laser field with Rabi frequency $\Omega $ interacts with the atoms
between the transition $|b\rangle \leftrightarrow |c\rangle $ with detuning $%
\Delta $. Our model is similar with quantum beat laser \cite{zubairy} except
with one movable mirror. The Hamiltonian of the hybrid system is given by

\begin{figure}[tbp]
\includegraphics[width=8.5cm,height=5cm]{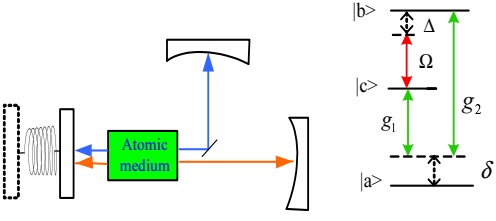}
\caption{(Color online) Sketch of the system and the atomic configuration.
Two-mode cavity fields interact with atomic transitions $|a\rangle
\leftrightarrow |c\rangle $ and $|a\rangle \leftrightarrow |b\rangle $ with
detuning $\protect\delta $, while the classical field drive the atomic level
between $|b\rangle \leftrightarrow |c\rangle $ with detuning $\Delta $.}
\end{figure}
\bigskip\ 
\begin{equation*}
H=H_{af}+H_{fo}+H_{dr}.
\end{equation*}%
The free energy of fields and the atoms as well as their interaction is

\begin{eqnarray}
H_{af} &=&\sum_{j=1,2}\omega _{j}\hat{a}_{j}^{\dagger }\hat{a}%
_{j}+\sum_{k=1}^{N}[\sum_{i=a,b,c}E_{i}\hat{\sigma}_{ii}^{(k)}+ \\
&&+(g_{1}\hat{a}_{1}\hat{\sigma}_{ca}^{(k)}+g_{2}\hat{a}_{2}\hat{\sigma}%
_{ba}^{(k)}+\Omega e^{-i\omega _{\Omega }t}\hat{\sigma}_{cb}^{(k)}+h.c.)], 
\notag
\end{eqnarray}%
The first term describes the energy of the two cavity modes with lowering
operator $a_{j}$ and cavity frequency $\omega _{j}$ (at equilibrium
position). The second term represents the energy of the atoms and the
interaction between the atoms and the cavity fields, where $\sigma
_{ij}=|i\rangle \langle j|$ is the spin operator of the atoms, $g_{j}$ is
the coupling between the cavity and the atoms, and $\Omega $ is the Rabi
frequency of the classical field driving the atoms between $|c\rangle $ and $%
|b\rangle $.

\begin{equation}
H_{fo}=\omega _{m}\hat{b}^{\dagger }\hat{b}+G(\hat{a}_{1}+\hat{a}%
_{2})^{\dagger }(\hat{a}_{1}+\hat{a}_{2})(\hat{b}+\hat{b}^{\dagger }),
\end{equation}%
where the first term is the free energy of the mechanical oscillator with
frequency $\omega _{m}$, and the second term represents the coupling between
the non-polarized two-mode fields and the mechanical resonator with
radiation-pressure coupling $G$, and the form of the coupling has been
employed to cool the mirror in \cite{cooling2}. We will show that the
coupling among the cavity fields and the resonator can enhance the
nonlinearity.

\begin{equation}
H_{dr}=\sum_{j=1,2}\varepsilon _{j}(\hat{a}_{j}^{\dagger }e^{-i\omega
_{L_{j}t}}+\hat{a}_{j}e^{i\omega _{L_{j}t}})
\end{equation}%
describes the two-mode cavity fields driven by weak classical fields.

Now, we switch into interaction picture rotating with $H_{0}=\sum_{j=1,2}%
\omega _{_{j}}\hat{a}_{j}^{\dagger }\hat{a}_{j}+\sum_{k=1}^{N}[%
\sum_{i=a,b,c}E_{i}\hat{\sigma}_{ii}^{(k)}+\delta \hat{\sigma}%
_{aa}^{(k)}-\Delta \hat{\sigma}_{cc}^{(k)}]$ where $\Delta
=E_{b}-E_{c}-\omega _{\Omega },$ $\delta =E_{c}-E_{a}-\omega
_{1}=E_{b}-E_{a}-\omega _{2}$. Then 
\begin{eqnarray}
H_{af1} &=&\sum_{k}[\Delta \hat{\sigma}_{cc}^{k}-\delta \hat{\sigma}%
_{aa}^{k}+(g_{1}\hat{a}_{1}\hat{\sigma}_{ca}^{k}+g_{2}\hat{a}_{2}\hat{\sigma}%
_{ba}^{k}  \notag \\
&&+\Omega \hat{\sigma}_{cb}^{k}+h.c.)],  \label{haf2}
\end{eqnarray}

\begin{equation}
H_{fo1}=\omega _{m}\hat{b}^{\dagger }\hat{b}+G(\hat{a}_{1}^{\dagger }\hat{a}%
_{1}+\hat{a}_{2}^{\dagger }\hat{a}_{2}+\hat{a}_{1}^{\dagger }\hat{a}%
_{2}e^{-idt}+\hat{a}_{2}^{\dagger }\hat{a}_{1}e^{idt})(\hat{b}+\hat{b}%
^{\dagger }),  \label{hfo}
\end{equation}

\begin{equation}
H_{dr1}=\sum_{j=1,2}\varepsilon _{j}(\hat{a}_{j}^{\dagger }+\hat{a}_{j}),
\label{hdr}
\end{equation}%
where $d=\omega _{2}-\omega _{1}$ is the frequency difference between the
two cavity modes. For simplicity, we have assume $\omega _{j}=\omega _{L_{j}}
$ ($j=1,2$). One can see that the Hamiltonian (\ref{hfo}) contains the terms 
$\hat{a}_{1}^{\dagger }\hat{a}_{2}\hat{b}+h.c.$ and $\hat{a}_{2}^{\dagger }%
\hat{a}_{1}\hat{b}+h.c.$ which are the typically nondegenerate parametric
amplification. That means, under certain range of parameters, one can obtain
squeezed states between one of the cavity modes and the mechanical
resonator. In addition, it had been shown in \cite{squeezing} that the
output field exhibits squeezing for single mode cavity optomechanics system
under strong drive condition. For the present coupling system, the squeezing
properties of the cavity fields deserve our further investigation using
linearized theory under strongly driven condition. But, we here focus on the
nonlinearity enhancement and show that the form of the coupling in (\ref{hfo}%
) plays an important role for enhancement of the nonlinearity.

Firstly, we consider large detuning condition and derive the effective
Hamiltonian of $H_{af1}$ (\ref{haf2}), and the motions for the atomic
operators $\hat{\sigma}_{ac}^{k}$ and $\hat{\sigma}_{ab}^{k}$ are given by 
\begin{eqnarray}
i\frac{d\hat{\sigma}_{ac}^{k}}{dt} &=&(\delta +\Delta )\hat{\sigma}%
_{ac}^{k}+g_{1}\hat{a}_{1}(\hat{\sigma}_{aa}^{k}-\hat{\sigma}_{cc}^{k})-g_{2}%
\hat{a}_{2}\hat{\sigma}_{bc}^{k}+\Omega \hat{\sigma}_{ab}^{k},  \notag \\
i\frac{d\hat{\sigma}_{ab}^{k}}{dt} &=&\delta \hat{\sigma}_{ab}^{k}-g_{1}\hat{%
a}_{1}\hat{\sigma}_{cb}^{k}+g_{2}\hat{a}_{2}(\hat{\sigma}_{aa}^{k}-\hat{%
\sigma}_{bb}^{k})+\Omega \hat{\sigma}_{ac}^{k}.  \label{elim}
\end{eqnarray}%
Under the large detuning conditions $\delta \gg \{g_{1},g_{2}\}$, $\Delta
\gg \Omega $, Eq. (\ref{elim}) can be solved adiabatically by taking ${d\hat{%
\sigma}_{ac}^{k}}/{dt}={d\hat{\sigma}_{ab}}/{dt}=0$. The adiabatic solutions
of $\hat{\sigma}_{ac}^{k}$ and $\hat{\sigma}_{bc}^{k}$ can then be
substituted into the Hamiltonian (\ref{haf2}). The most of the atoms are in
their ground state $|a\rangle $, thus, by elimination of the atomic
variables the effective Hamiltonian describing the interaction between the
two mode fields can be written as%
\begin{eqnarray}
H_{af2} &=&-\nu _{1}\hat{a}_{1}^{\dag }\hat{a}_{1}-\nu _{2}\hat{a}_{2}^{\dag
}\hat{a}_{2}  \notag \\
&&+\lambda (\hat{a}_{1}^{\dag }\hat{a}_{2}+\hat{a}_{2}^{\dag }\hat{a}_{1}),
\label{haff}
\end{eqnarray}

where%
\begin{eqnarray*}
\nu _{1} &=&\frac{2g_{1}^{2}\delta N}{\tilde{\Delta}},\nu _{2}=\frac{%
2g_{2}^{2}(\delta +\Delta )N}{\tilde{\Delta}}, \\
\lambda  &=&\frac{2g_{1}g_{2}\Omega N}{\tilde{\Delta}},
\end{eqnarray*}%
with $\tilde{\Delta}=\delta (\Delta +\delta )-\Omega ^{2}$. We see that an
additional interaction term between the two-mode cavity fields is introduced
because of the interaction between atomic media and the fields. As shown in
Fig.1, the two-mode fields interact with the atoms between the transitions $%
|a\rangle $ $\leftrightarrow $ $|b\rangle $ and $|a\rangle $ $%
\leftrightarrow $$|c\rangle $ (similar with V-type configuration). Even
though most of the atoms are in their ground state, the participation of the
atoms still induce a beam splitter type interaction shown in Eq.(\ref{haff}%
). It is the term proportional to the number of atoms that enhances the
cross-Kerr nonlinearity.

Although we study the large detuning case, the value of $\nu _{i}$ and $%
\lambda $ can be large, because it is proportional to the number of atoms;
therefore we can switch into a picture rotating with $\omega _{m}\hat{b}%
^{\dagger }\hat{b}+$ $H_{af2}$ and treat the other terms as perturbation. In
order to do that, we diagonalize the Hamiltonian (\ref{haff}) by defining $%
\hat{c}_{1}=\hat{a}_{1}\cos \theta +\hat{a}_{2}\sin \theta $ and $\hat{c}%
_{2}=\hat{a}_{1}\sin \theta -\hat{a}_{2}\cos \theta $, and then we have 
\begin{equation}
H_{af3}=-\omega _{c_{1}}\hat{c}_{1}^{\dag }\hat{c}_{1}-\omega _{c_{2}}\hat{c}%
_{2}^{\dag }\hat{c}_{2},  \label{haff2}
\end{equation}%
with%
\begin{eqnarray}
\omega _{c_{1}} &=&\nu _{1}\cos ^{2}\theta +\nu _{2}\sin ^{2}\theta -\lambda
\sin 2\theta ,  \label{theta} \\
\omega _{c_{2}} &=&\nu _{1}\sin ^{2}\theta +\nu _{2}\cos ^{2}\theta +\lambda
\sin 2\theta ,  \notag \\
tg2\theta  &=&\frac{2\lambda }{\nu _{2}-\nu _{1}}.  \notag
\end{eqnarray}%
Now we jointly consider Hamiltonian (\ref{hfo}) and (\ref{haff2}) and switch
into a picture rotating with $H_{0}^{\prime }=\omega _{m}\hat{b}^{\dagger }%
\hat{b}-\omega _{c_{1}}\hat{c}_{1}^{\dag }\hat{c}_{1}-\omega _{c_{2}}\hat{c}%
_{2}^{\dag }\hat{c}_{2}$. Then, we use the effective Hamiltonian method
proposed in \cite{effective} and have the effective Hamiltonian as 
\begin{eqnarray}
H_{eff} &=&\eta _{1}\hat{c}_{1}^{\dag }\hat{c}_{1}\hat{c}_{2}^{\dag }\hat{c}%
_{2}+\eta _{2}(\hat{c}_{1}^{\dag }\hat{c}_{1}-\hat{c}_{2}^{\dag }\hat{c}_{2})%
\hat{b}^{\dagger }\hat{b}+  \label{heff} \\
&&+s[(\hat{c}_{1}^{\dag }\hat{c}_{1})^{2}+(\hat{c}_{2}^{\dag }\hat{c}%
_{2})^{2}]+u_{1}\hat{c}_{1}^{\dag }\hat{c}_{1}+u_{2}\hat{c}_{2}^{\dag }\hat{c%
}_{2},  \notag
\end{eqnarray}%
where%
\begin{eqnarray}
\eta _{1} &=&v+u_{2}-u_{1},\eta _{2}=u_{2}-u_{1},  \label{hco} \\
v &=&[\frac{\omega _{m}\sin ^{2}2\theta }{\omega _{m}^{2}-d^{2}}-\frac{2}{%
\omega _{m}}]G^{2},  \notag \\
u_{1} &=&\frac{G^{2}\sin ^{4}\theta }{\omega _{f}+d-\omega _{m}}+\frac{%
G^{2}\cos ^{4}\theta }{\omega _{f}-d-\omega _{m}},  \notag \\
u_{2} &=&-\frac{G^{2}\sin ^{4}\theta }{\omega _{f}+d+\omega _{m}}-\frac{%
G^{2}\cos ^{4}\theta }{\omega _{f}-d+\omega _{m}},  \notag \\
s &=&-[\frac{1}{\omega _{m}}+\frac{\omega _{m}\sin ^{2}2\theta }{2\omega
_{m}^{2}-2d^{2}}]G^{2},  \notag \\
\omega _{f} &=&\omega _{c2}-\omega _{c_{1}}.  \notag
\end{eqnarray}%
The first term of the effective Hamiltonian (\ref{heff}) describes the
cross-Kerr nonlinearity between the two cavity modes, and the second term is
that between one of the cavity modes and the oscillator. The third term is
the Kerr nonlinearity of cavity fields. 
\begin{figure}[tph]
\includegraphics[width=8.5cm,height=5cm]{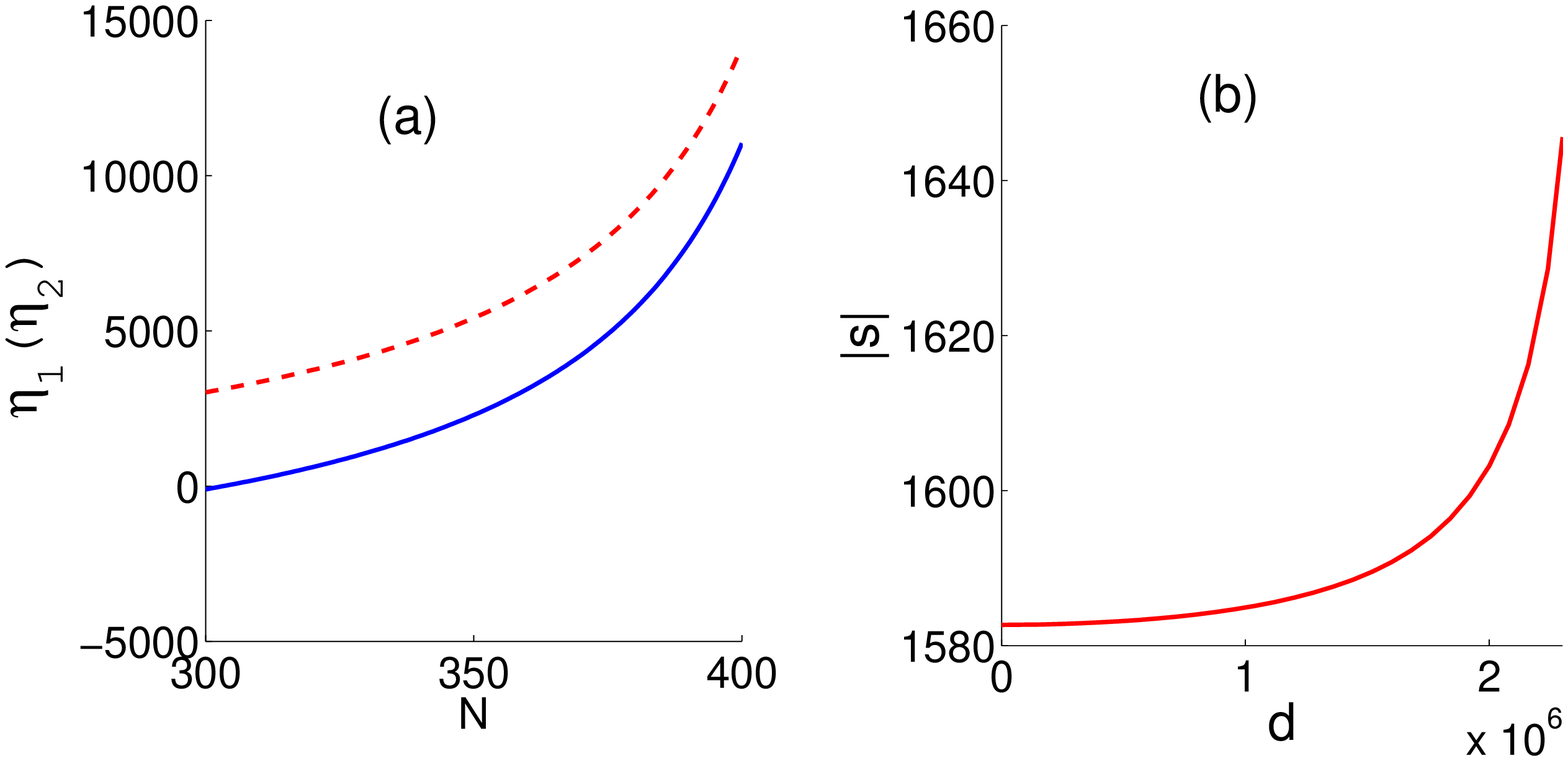}
\caption{(Color online) The enhancement of the strength of the cross-Kerr
and the Kerr nonlinearity. (a): $\protect\eta _{1}$($\protect\eta _{2})$ is
proportional to the number of atoms where $d=200\protect\pi KHz$. (b): $|s|$
change with $d$ \ for $N=350$. The other parameters are $g_{1}=20\protect\pi %
KHz$, $g_{2}=22\protect\pi KHz$, $\protect\omega _{m}=800\protect\pi KHz$, $%
G=20\protect\pi KHz$, $\Omega =20\protect\pi KHz$, $\protect\delta =\Delta
=250\protect\pi KHz$. }
\end{figure}

In Fig.2a, we show that the cross-Kerr nonlinearity strength $\eta _{1}$ and 
$\eta _{2}$ increase with the increasing of the number of atoms at certain
range of parameters, which means that the cross-nonlinearity can be enhanced
by increasing of the number of atoms. From the coefficients of expression (%
\ref{hco}), we know that the reason of $\eta _{i}$ changing with the number
of atoms lies in the frequency difference $\omega _{f}$ between quasi-mode $%
\omega _{c_{1}}$ and $\omega _{c2}$ modulated by the number of atoms $N$.
Usually, it is difficult to adjust the radiation pressure coupling $G$,
because it is determined by the mirror and the cavity. Therefore, to enhance
the cross-Kerr nonlinearity, we should decrease $|\omega _{f}\pm d\pm \omega
_{m}|$. If $|\omega _{f}\pm d\pm \omega _{m}|<\omega _{m}$ , the
nonlinearity is larger than $\frac{G^{2}}{\omega _{m}^{2}}$ \cite{gongzr}.
The frequency difference $d$ between two modes and the frequency difference $%
\omega _{f}$ between quasi-mode $\omega _{c_{1}}$ and $\omega _{c2}$ can
favor us do that. We can achieve our target by controlling the number of
atoms and keep an appropriate value of $d$. Of course, in the above process,
we should keep \{$Gsin^{2}\theta $,$G\cos ^{2}\theta \}<|\omega _{f}\pm d\pm
\omega _{m}|$ so as to meet the condition of effective Hamiltonian
approximation. Comparing with \cite{max} where the tunneling rate between
two cavities is the key factor to enhance the nonlinearity, the modulation
of the nonlinearity by adjusting the number of atoms is easier to implement
and control. Fig.2b shows us the strength of Kerr nonlinearity $|s|$ (in
unit of Hz) as a function of $d$. Although $s$ is not influenced by $N$ ($%
\theta $ has nothing to do with $N$ see Eq.(\ref{theta})), the nonlinearity $%
|s|$ can be enhanced, because one of the denominators of $s$ is decreased by 
$d$ while this behavior can not exist just in single cavity optomechanical
system \cite{gongzr}. Furthermore, because the photon-blocked effect of the
third term, it is easy to perform QND of single photon, while this property
has no exhibition in \cite{max}.

\section{The quantum-nondemolition measurement of photon}

Cross-Kerr nonlinearity is believed as high efficiency quantum-nondemolition
measurement \cite{nondemo,nonde3}. It also can be used to perform quantum
gate \cite{gate1,gate2}, quantum information procession \cite{inf,inf2}, and
entanglement generation \cite{ent1,ent2,ent3,ent4}. As a usage of the
present scheme, we now study the QND measurement of phonon and photon.

The master equation of the system is 
\begin{equation}
\frac{d\hat{\rho}}{dt}=-i[\hat{H}_{I},\rho ]+\sum_{i=1}^{4}D[\hat{A}_{i}]%
\hat{\rho},
\end{equation}%
where $\hat{H}_{I}=\hat{H}_{dr1}+\hat{H}_{af2}+\hat{H}_{fo1}$, $D[\hat{A}%
_{i}]=\hat{A}_{i}\rho A_{i}^{\dagger }-\frac{1}{2}\rho A_{i}^{\dagger }\hat{A%
}_{i}-\frac{1}{2}\hat{A}_{i}^{\dagger }\hat{A}_{i}\rho $, $\hat{A}_{1}=\sqrt{%
2\kappa _{1}}a_{1},\hat{A}_{2}=\sqrt{2\kappa _{2}}a_{2}$, $\hat{A}_{3}=\sqrt{%
2\gamma _{m}(n_{th}+1)}\hat{b}$, and $\hat{A}_{4}=\sqrt{2\gamma _{m}n_{th}}%
\hat{b}^{\dagger }$. In order to illustrate the nonlinearity, we employ the
Hamiltonian in the interaction picture rather than using effective
Hamiltonian (11). Due to the high frequency of the cavity and the large
frequency difference between the cavity fields and the movable mirror, the
environment of the cavity fields can be treated as zero temperature while
the mirror should be in thermal field. 
\begin{figure}[tbph]
\includegraphics[width=8.cm,height=7.5cm]{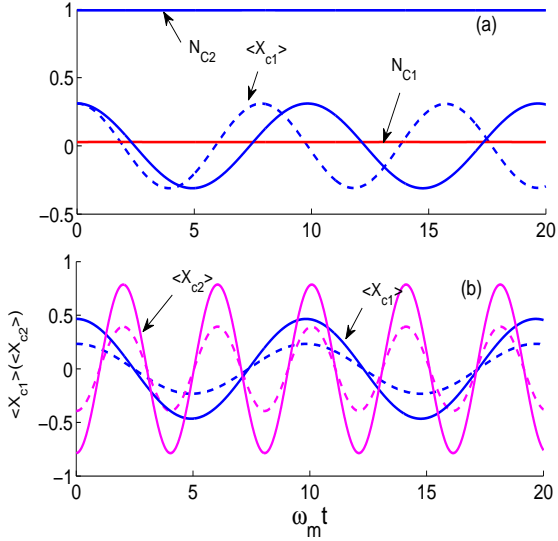}
\caption{(Color online) The existence of cross-Kerr nonlinearity via
quadrature $\langle X\rangle $ measurement. (a): The evolution of $\langle
X_{C_{1}}\rangle $ for initial state $|0.1_{c_{1}},1_{c_{2}}\rangle $, with $%
N=320$ (solid line) and $400$ (dashed line). (b): $\langle X_{C_{1}}\rangle $
and $\langle X_{C_{2}}\rangle $ for two initial coherent states $%
|0.1_{c_{1}},0.2_{c_{2}}\rangle $ (dashed)and $|0.2_{c_{1}},0.4_{c_{2}}%
\rangle $ \ (solid line) where $N=380$. The other parameters are the same
with Fig.2.}
\end{figure}

Now we first illustrate the existence of cross-Kerr nonlinearity within the
system. For simplicity, we omit the pumping fields and the loss of the
fields and assume that the oscillator is in its ground state. In Fig.3a, $%
c_{1}$ is in coherent state $|\alpha \rangle $ $(\alpha =0.1)$ , and the
mode $c_{2}$ is initially in Fock state $|1\rangle $. If the system contains
the cross- Kerr nonlinearity with the term $\eta _{1}\hat{c}_{1}^{\dag }\hat{%
c}_{1}\hat{c}_{2}^{\dag }\hat{c}_{2}$, then we will have $e^{-i\eta _{1}\hat{%
c}_{1}^{\dag }\hat{c}_{1}\hat{c}_{2}^{\dag }\hat{c}_{2}t}|\alpha ,1\rangle
_{c_{1},c_{2}}=|\alpha e^{i\eta _{1}t},1\rangle $ , the $c_{1}$ mode
acquires a phase. When the phase equals to $\pi $, a two-photon
controlled-phase gate is naturally implemented, from which a CNOT gate can
also be easily constructed \cite{gate1,gate2}. Performing quadrature
operator $x=c_{1}+c_{1}^{\dagger }$ measurement by the homodyne apparatus,
we know $\langle x\rangle =2\alpha \cos \eta _{1}t$ . Employing the
Hamiltonian $\hat{H}=\hat{H}_{aff1}+\hat{H}_{fo2}$, we plot $\langle \hat{c}%
_{1}^{\dag }\hat{c}_{1}\rangle $ ,$\langle \hat{c}_{2}^{\dag }\hat{c}%
_{2}\rangle $, $\langle X_{c_{1}}\rangle $ and $\langle X_{c_{2}}\rangle $
in Fig.3. We see that the photon number almost keeps unchanged but $\langle
x\rangle $ oscillates with cosine function. In addition, the frequency of
cosine function increases with the increasing of $N$ (see Fig.3a), that is
to say, the strength of the cross-Kerr nonlinearity enhances with the
increasing of $N$. For both of the quasi-mode initially in coherent state,
Fig.3b shows that the periods of the oscillation of $\langle
X_{c_{1}}\rangle $ and $\langle X_{c_{2}}\rangle $ have nothing to do with
different initial coherent state, only the amplitudes are affected by the
coherent states. Although we plot the figure from $\hat{H}=\hat{H}_{aff1}+%
\hat{H}_{fo2}$, one can clearly see that cross-Kerr nonlinearity do exist in
the two-mode optomechanical system. 
\begin{figure}[tbph]
\includegraphics[width=7.cm,height=7.5cm]{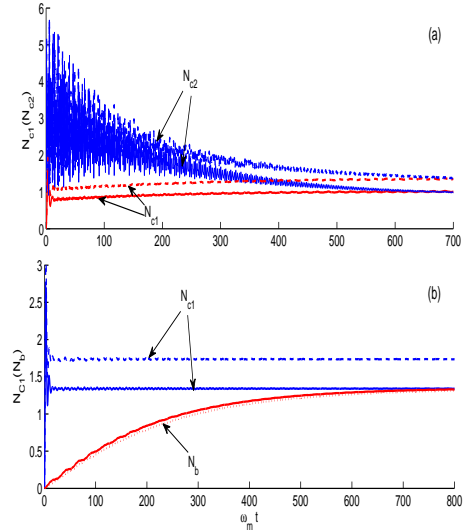}
\caption{(Color online) The QND of photon number as well as phonon number.
The QND of $N_{C_{2}}$ via detection of $N_{C_{1}}$ (a) where $\protect%
\varepsilon _{1}=23g_{1}$, $\protect\varepsilon _{2}=62g_{1}$, and the QND
of phonon number $N_{b}$ via detection of $N_{C_{1}}$ (b) where $\protect%
\varepsilon _{1}=31g_{1}$, $\protect\varepsilon _{2}$=$\protect\varepsilon %
_{1}\tan \protect\theta $, $n_{th}=4$. For both of the figures $\protect%
\kappa _{1}=200\protect\pi KHz$, $\protect\kappa _{2}=0.01\protect\kappa _{1}
$, $\protect\gamma _{m}=0.001\protect\kappa _{1}$, $N=320$ (dashed line), $%
380$ (solid line), and the other parameters are the same with Fig.2.}
\end{figure}

We now discuss the QND of photon and phonon number via detecting the photon
number of $c_{1}$. In Fig.4a, initially there is a single photon in mode $%
c_{2}$, $c_{1}$ is in vacuum state, and the phonon is in thermal state with $%
T=42\mu K$. In order to perform QND in mode $c_{2}$, we need a weak
classical driving field to compensate the loss of the cavity so as to keep
the single photon, and we also need a little bit strong driving field to
pump mode $c_{1}$. Because of the detection of mode $c_{1}$, we have assumed
that the loss of mode $c_{1}$ is larger than that of $c_{2}$ and phonon
mode. For the case $cos\theta \approx 1$ in the present parameter region,
the above condition by assuming $\kappa _{1}\gg \kappa _{2}$ can be
realized. Fig.4a clearly shows us that after a period of time evolution, the
photon number of mode $c_{1}$ and $c_{2}$ are equal, which means that the
leaking-out photon number of mode $c_{1}$ is exactly equal to that of mode $%
c_{2}$; thus we can perform QND of $c_{2}$ by detecting $c_{1}$. If we want
to detect phonon number, we need the phonon excitations which can be
obtained by employing the thermal field. Fig.4b displays that under
appropriate condition, the QND of phonon number also can be performed by
detecting the same mode $c_{1}$, where we assume that the mode $c_{1}$ and $%
c_{2}$ are both initially in vacuum state, the phonon mode is in thermal
state with $T=42\mu K$, and the mode $c_{2}$ is not driven under the
condition $\varepsilon _{2}=\varepsilon _{1}\tan \theta $. From Fig.4, we
conclude that two initial independent bosonic modes finally acquire
identical intensity after evolution, which is the so called synchronization.
In addition, the number of atoms does affect the behavior of
synchronization. For less number of atoms, the time of evolution to achieve
synchronization is longer than that with more atoms. Comparing dashed line ($%
N=320$) with solid line ($N=380$), we can clearly observe it. Thus, the
number of atoms do modulate the nonlinearity. We can enhance the nonlinear
strength by increasing the number of atoms in the present region of
parameters. As we know that nonlinear interaction is necessary for the
emergence of synchronization \cite{syn1,mari2}; therefore, we can safely say
that the cross-nonlinearity within the present scheme really can offer us
QND photon and phonon numbers.

During the review process of the paper, we read the related works \cite%
{Brkj,Lemonde}. For strongly driven optomechanical system, they included the
nonlinear interaction term usually omitted by most works and shown that
intrinsic nonlinear are observable even with a relatively weak
optomechanical coupling. Different from \cite{Brkj,Lemonde}, we discuss the
weakly driven and weak coupling condition, we do not employ the so called
linearized optomechanics. Most importantly, we put forward an alternatively
scheme to enhance the cross-Kerr nonlinearity via atomic coherence.

\section{Conclusion}

We put forward a scheme employing atomic coherence to enhance the
nonlinearity of optomechanical system. When the atoms interact with the two
mode fields with large detuning condition, we adiabatically eliminated the
degree of atoms. For weak coupling among the two mode fields and the
mechanical resonator, we derived the effective Hamiltonian and shown that
the nonlinear coefficients can be enhanced by increasing the number of atoms
at certain range of parameters. We also numerically studied the nonlinearity
enhancement beyond the effective Hamiltonian. Furthermore, we investigated
the potential quantum nondemolition measurement of bosonic mode. Our results
show that the present system exhibits synchronization, and the nonlinear
effects provide us a means in performing QND.

As to the realizability in experiment, we do not demand the strong coupling
of atoms with the cavity, for example, our parameters satisfy $g_{1}\ll
\kappa _{1}$, $g_{2}\approx \kappa _{2}$. We also do not require strong
coupling between the cavity fields and the mechanical resonator, and one can
easily check that the so called strong coupling condition $G^{2}>\kappa
\omega _{m}$ is not meet. Although the above strong coupling conditions are
achievable, the loosening of the coupling conditions is more realizable and
is still valuable.

\begin{acknowledgments}
Acknowledgments: Ling Zhou,Jiong Cheng and Yan Han are supported by NSFC
(Grant Nos.11074028). Weiping Zhang acknowledges the support of NSFC (Grant
Nos. 10828408, 10588402), the National Basic Research Program of China (973
Program) under Grant No. 2011CB921604. All authors thank Open Fund of the
State Key Laboratory of Precision Spectroscopy, ECNU.
\end{acknowledgments}

\end{document}